\documentclass[11pt]{cernrep}
\usepackage{graphicx}
\usepackage{here}
\begin{document}
 \title{FINITE TEMPERATURE FIELD THEORIES ON THE LIGHT-FRONT}
\author{Ashok Das}
\institute{Department of Physics and Astronomy, University of
  Rochester, Rochester, NY 14627, USA.}
\maketitle
\begin{abstract}
In this talk, we describe our present understanding of thermal
field theories on the light-front with an application to Schwinger
model. 
\end{abstract}

\section{INTRODUCTION}

Light-front field theories [1] have been of great interest for a
variety of reasons and more recently, there have been a lot of
interest in the study of such theories at finite temperature
[2,3,4]. At first sight, the study of this question would appear to be
quite straightforward. Namely, since in the light-front formalism,
\begin{equation}
p_{+} = \frac{1}{\sqrt{2}} (p_{0}+p_{3})\, ,
\end{equation}
plays the role of the Hamiltonian, it would seem natural to define the
finite temperature partition function of the system to correspond to
\begin{equation}
Z (\beta ) = {\rm Tr}\,e^{\beta p_{+}}\, ,
\end{equation}
where $\beta$ represents the inverse temperature in units of the
Boltzmann constant. As we will see, such a generalization leads to
problems  and in this talk we would discuss an alternate generalization
of the partition function as well as applications following from it.

The organization of the presentation is as follows. In section {\bf
  2}, we would discuss the problems associated with the naive
  generalization of the partition function in (2) through a simple
  example and propose an alternate generalization of the partition
  function for light-front field theories. In section {\bf 3}, we will
  describe the systematic derivation of such a generalization. In
  section {\bf 4}, we will apply the alternate definition of the
  partition function to study various questions at finite temperature
  associated with the Schwinger model such as the anomaly, condensates
  and bound states. We will close with a short conclusion in section
  {\bf 5}.

\section{DIFFICULTIES WITH THE NAIVE GENERALIZATION}

Let us, for completeness, note that in the light-front formalism, one
defines 
\begin{equation}
x^{\pm} = \frac{1}{\sqrt{2}} (x^{0} \pm x^{3})\, ,
\end{equation}
where $x^{+}$ is identified with the time coordinate, $x^{-}$
the longitudinal coordinate, while the transverse coordinates
remain unchanged. Correspondingly, we have
\begin{equation}
p_{\pm} = \frac{1}{\sqrt{2}} (p_{0}\pm p_{3})\, ,
\end{equation}
where $p_{+}$ can be identified with energy (Hamiltonian). In the
light-front coordinates, the metric is off-diagonal with the form
\begin{equation}
\eta^{+-} = \eta^{-+} = 1,\quad \eta^{ij} = -\delta^{ij},\quad
\eta^{++}=\eta^{--} =0,\quad i,j = 1,2\, .
\end{equation}

We note that since conventionally, for a system with Hamiltonian $H$,
the density matrix is given by
\begin{equation}
\rho (\beta) = e^{-\beta H}\, ,
\end{equation}
it is natural to consider the density matrix for light-front field theories
to be given by
\begin{equation}
\rho (\beta_{\rm LF}) = e^{-\beta_{\rm LF} p_{+}} = e^{-\beta_{\rm LF}
  p^{-}}\, ,
\end{equation}
where $\beta_{\rm LF}$ denotes the inverse of the temperature in the
light-front frame. We will see in the next section how to identify
this temperature, but for the moment, let us assume for simplicity
that
\begin{equation}
\beta_{\rm LF} = \beta\, .
\end{equation}

Using Eqs. (7) and (8), we can now define the propagators in any
theory. For example, for a massive scalar theory the propagagtor in the
imaginary time formalism will have the form
\begin{equation}
G (p) = \frac{1}{2p^{-}p^{+} - \omega_{p}^{2}}\, ,
\end{equation}
where $p^{-} = 2i\pi nT$ and $\omega_{p}^{2} = {\bf p}^{2} +
m^{2}$ with $n$ an integer. On the other hand, in the real time
formalism, the relevant component of the propagator (from one loop
point of view) would have the form
\begin{equation}
G_{++} (p) = \frac{1}{2p^{-}p^{+} - \omega_{p}^{2}} - 2i\pi n_{\rm B}
(|p^{-}|) \delta (2p^{-}p^{+} - \omega_{p}^{2})\, ,
\end{equation}
where $n_{\rm B}$ denotes the bosonic distribution function
\begin{equation}
n_{\rm B} (|p^{-}|) = \frac{1}{e^{\beta |p^{-}|} - 1}\, .
\end{equation}

The sign of trouble in the naive generalization of the density matrix
(and, therefore, the partition function) is already manifest in
(11). Namely, the distribution function is not damped at
$p^{-}\rightarrow 0$. We are, of course, used to such a divergent
behavior in massless theories, but here we are considering a massive
scalar theory.

To see further signs of trouble, let us calculate the one loop
self-energy  in the $\phi^{4}$ theory in $D$ dimensions. The
finite  temperature
part is easily analyzed in the real time formalism and has the form
\begin{equation}
i\Pi_{++}^{(\beta)} (p) = - \frac{i\lambda}{2(2\pi)^{D-1}} \int
\mathrm{d}^{D}k\, 
n_{B} (|k^{-}|) \delta (2k^{-}k^{+} - \omega_{k}^{2})\, .
\end{equation}
We note the peculiar feature in (12) that under the redefinition
\begin{equation}
k^{-}\rightarrow \frac{k^{-}}{\beta},\qquad k^{+}\rightarrow \beta
k^{+}\, ,
\end{equation}
the integral becomes independent of temperature, even though this is
the finite temperature part of the self-energy. This is again a signal
of the divergent structure of the integrand. In fact, one can evaluate
this integral in a regulated manner as
\begin{eqnarray}
i\Pi_{++}^{(\beta)} (p) & = & -\frac{i\lambda}{2(2\pi)^{D-1}} \int
\mathrm{d}^{D-2} \int_{0}^{\infty}
\frac{\mathrm{d}k^{+}}{k^{+}}\,n_{\rm B}
\left(\frac{\omega_{k}^{2}}{2k^{+}}\right)\nonumber\\
\noalign{\vskip 4pt}%
 & = & \lim_{\epsilon\rightarrow 0}\,- \frac{i\lambda}{2 (2\pi)^{D-1}}
\int \mathrm{d}^{D-2}k \int_{0}^{\infty}
\frac{\mathrm{d}k^{+}}{(k^{+})^{1+\epsilon}}\, n_{B}
\left(\frac{\omega_{k}^{2}}{2k^{+}}\right)\nonumber\\
\noalign{\vskip 4pt}%
 & = & \lim_{\epsilon\rightarrow 0}\,-\frac{i\lambda}{2(2\pi)^{D-2}}
\int \mathrm{d}^{D-2}k \left(\frac{1}{\epsilon} + \ln \frac{2}{\beta
  \omega_{k}^{2}} + {\rm constant}\right)\, .
\end{eqnarray}
In deriving this result, we have set a reference mass scale to unity
for convenience.

There are several things to note from the form of the amplitude in
(14). First, it is exact in the sense that we have not made any high
or low $T$ expansion. Second, it is divergent and the divergence
structure is independent of temperature. This is a particularly
serious problem in that it would suggest that one would need new zero
temperature counterterms at finite temperature. This is completely
counter intuitive in that the interaction with a thermal medium takes
place on-shell (which is rather clear in the real time formalism of
(10)) and would render the question of renormalization intractable in
such theories. We note that the temperature dependence of the
amplitude has the same logarithmic behavior independent of the number
of dimensions which is not consistent with our experience [5] in
thermal field theories. Note also that the temperature dependent part
of the amplitude in (14) diverges for any $D$ other than $D=2$ thereby
necessitating finite temperature counterterms. Finaly, let us note
that in the limit of zero temperature ($\beta\rightarrow \infty$),
this amplitude does not vanish even though this is supposed to
represent the  finite temperature part.

The source of the problem is not hard to see. The density matrix in
(6) is defined in a frame where the heat bath is at rest. On the other
hand, on the light-front, we cannot have a heat bath at
rest. Consequently, our starting point for generalization is not quite
correct. An alternate generalization for the density matrix can be
given as [2]
\begin{equation}
\rho (\beta) = e^{-\beta \frac{p^{+}+p^{-}}{\sqrt{2}}} = e^{-\beta
  p^{0}}\, .
\end{equation}
The physical basis for such a generalization will be discussed in the
next section. For the moment, let us note that it seems like we are
dealing with 
the conventional density matrix, but one must remember that, in
light-front field theories, the time ordering in correlation functions
is with respect to $x^{+}$.

With the alternative generalization, the propagator for the massive
scalar theory will have the form (in imaginary time formalism)
\begin{equation}
G (p) = \frac{1}{2p^{-}p^{+} - \omega_{p}^{2}}\, ,
\end{equation}
with $p^{-} = 2\sqrt{2}i\pi n T - p^{+}$. The corresponding component
of the real time propagator will have the form
\begin{equation}
G_{++} (p) = \frac{1}{2p^{-}p^{+} - \omega_{p}^{2}} - 2i\pi n_{\rm B}
\left(\left|\frac{p^{-}+p^{+}}{\sqrt{2}}\right|\right) \delta
(2p^{-}p^{+} - \omega_{p}^{2})\, .
\end{equation}
We note that the distribution function, in this case, has proper
damping. We can now evaluate the one-loop scalar self-energy
in the $\phi^{4}$ theory and the thermal part of the amplitude now can
be evaluated in the high temperature limit (in $D=4$) to have the
behavior
\begin{eqnarray}
i\Pi_{++}^{(\beta)} (p) & = & - \frac{i\lambda}{2 (2\pi)^{3}} \int
\mathrm{d}^{2}k \int_{0}^{\infty} \frac{\mathrm{d}k^{+}}{k^{+}}\,
n_{\rm B} \left(\frac{\omega_{k}^{2} + 2 (k^{+})^{2}}{2\sqrt{2}
  k^{+}}\right)\nonumber\\
\noalign{\vskip 4pt}%
 & \approx & - \frac{i\lambda}{24 \beta^{2}}\, ,
\end{eqnarray}
which is well behaved and coincides with the result obtained in the
conventional  thermal field theories [5]. The thermal self-energy can
also be calculated in the $\phi^{3}$ theory [2] and although at first
sight, the result looks quite different from the conventional one,
Weldon has shown [3] through a clever change of coordinates that they
are indeed identical.

\section{A SYSTEMATIC STUDY}

In this section, we will describe following [3] a systematic way of
studying thermal field theories in different frames, which will also
clarify the meaning of the density matrix in (15).

Let us consider a physical system in a generalized coordinate system
related to the conventional one through a linear invertible
transformation of the form
\begin{equation}
\bar{x}^{\mu} = L^{\mu}_{\,\alpha} x^{\alpha},\qquad x^{\alpha} =
L^{\alpha}_{\,\mu} \bar{x}^{\mu}\, .
\end{equation}
It follows from (19) that
\begin{equation}
L^{\mu}_{\,\alpha} L^{\alpha}_{\,\nu} = \delta^{\mu}_{\nu},\qquad
L^{\alpha}_{\,\mu} L^{\mu}_{\,\beta} = \delta^{\alpha}_{\beta}\, .
\end{equation}
An arbitrary vector will transform under this redefinition as
\begin{equation}
\bar{V}^{\mu} = L^{\mu}_{\,\alpha} V^{\alpha},\qquad \bar{V}_{\mu} =
V_{\alpha} L^{\alpha}_{\,\mu}\, .
\end{equation}
It follows from (20) and (21) that a scalar will remain invariant
under such a redefinition while the metric tensor will transform as
\begin{equation}
\bar{g}^{\mu\nu} = L^{\mu}_{\,\alpha} L^{\nu}_{\,\beta}
g^{\alpha\beta},\qquad \bar{g}_{\mu\nu} = L^{\alpha}_{\,\mu}
L^{\beta}_{\,\nu} g_{\alpha\beta},\qquad \bar{g}^{\mu\lambda}
\bar{g}_{\lambda\nu} = \delta^{\mu}_{\nu}\, .
\end{equation}
It is worth emphasizing here that such a coordinate redefinition does
not necessarily represent a Lorentz transformation and as a result,
the form of the metric can be different in different frames. However,
the volume element will remain the same in the two frames, namely,
\begin{equation}
\sqrt{-g} \mathrm{d}^{4} x = \sqrt{-\bar{g}} \mathrm{d}^{4}\bar{x}\, .
\end{equation}

Let us next see how we can describe statistical mechanics in the new
coordinate system. First, let us recall that when we have a system
interacting with a heat bath which is moving with a normalized
velocity $u^{\alpha}$ (with $u^{\alpha}u_{\alpha} = 1$), the density
matrix is given by
\begin{equation}
\rho (\beta) = e^{-\beta u^{\alpha}p_{\alpha}}\, .
\end{equation}
In the transformed coordinate (since scalars do not transform), the
density matrix, therefore, will have the form
\begin{equation}
\rho (\beta) = e^{-\beta \bar{u}^{\mu}\bar{p}_{\mu}}\, .
\end{equation}
In the rest frame of the heat bath in the transformed coordinates, the
four velocity will have the form
\begin{equation}
\bar{u}^{\mu}_{\rm rest} = \left(\frac{1}{\sqrt{\bar{g}_{00}}}, 0, 0,
0\right)\, ,
\end{equation}
so that in the rest frame of the heat bath in the transformed
coordinates, the density matrix will have the form
\begin{equation}
\rho = e^{-\frac{\beta}{\sqrt{\bar{g}_{00}}}\bar{p}_{0}} =
e^{-\bar{\beta} \bar{p}_{0}}\, .
\end{equation}
This identifies the inverse temperature in the transformed coordinate
system to be
\begin{equation}
\bar{\beta} = \frac{\beta}{\sqrt{\bar{g}_{00}}}\, .
\end{equation}
The density matrix can, in fact, be checked to satisfy the KMS
condition with the periodicity given by $\bar{\beta}$.

Let us now specialize to light-front field theories. In this case, the
transformations can be easily constructed, but most important is the
observation that in these coordinates, the metric has the form
(suppressing the transverse degrees of freedom) as given in (5),
namely,
\begin{equation}
\bar{g}_{\mu\nu} = \left(\begin{array}{cc}
0 & 1\\
\noalign{\vskip 4pt}%
1 & 0
\end{array}\right)\, ,
\end{equation}
so that, in this case, we have $\bar{g}_{00} = 0$.  As a result, 
statistical description fails for conventional light-front theories in
the sense that any finite temperature
is mapped to zero temperature. The reason for this failure, as we have
noted earlier, can be seen from (26) to be the absence of a rest frame
for the heat bath.

To proceed further, let us note that light-front quantization only
requires that we identify $x^{+}$ with time and quantize on equal
$x^{+}$ surfaces. It does not say anything about how the other
coordinates should be chosen. We can try to take advantage of this and
see if a suitable light-front coordinate system can be defined where
statistical description is possible. we note that conventional light-front
quantization has several attarctive features (such as the linearity of
dispersion relation, the large number of kinematic generators etc) and
choosing a new coordinate system, it would be desirable to maintain
such features as much as is possible. Taking into account various
considerations (as well as the desire to identify $\beta_{\rm LF} =
\beta$), it turns out that the redefinition
\begin{equation}
\bar{x}^{0} = x^{0}+x^{3},\qquad \bar{x}^{3} = x^{3}\, ,
\end{equation}
has all the desired properties and yet allows a statistical
description. we note that under the redefinition,
\begin{equation}
\bar{p}_{0} = p_{0},\qquad \bar{p}_{3} = - p_{0}+p_{3}\, ,
\end{equation}
and the transformed metric has the form (suppressing the transverse
degrees) 
\begin{equation}
\bar{g}_{\mu\nu} = \left(\begin{array}{rr}
1 & -1\\
-1 & 0
\end{array}\right),\qquad \sqrt{-\bar{g}} = 1\, .
\end{equation}
As a result, the density matrix has the form
\begin{equation}
\rho (\beta ) = e^{-\beta \bar{p}_{0}} = e^{-\beta p_{0}}\, ,
\end{equation}
which can be compared with the alternate generalization in (15). 

\section{SCHWINGER MODEL}

With the density matrix in (15) (or (33)), one can now define the
propagators for various theories both in the imaginary time as well as
the real time formalisms [4] and can carry out calculations. In this
section, we will discuss some of the results obtained from an
application of this formalism to light-front Schwinger model at finite
temperature. 

The Schwinger model describes massless QED in $1+1$ dimensions. Under
a redefinition of coordinates in (30), it is straightforward to see
that the Dirac matrices transform as
\begin{equation}
\bar{\gamma}^{0} = \gamma^{0} + \gamma^{1},\qquad \bar{\gamma}^{1} =
\gamma^{1}\, .
\end{equation}
In terms of the transformed Dirac matrices, let us define the
projection operators,
\begin{equation}
P^{-} = - \frac{1}{2} \bar{\gamma}^{0}\bar{\gamma}^{1} = \frac{1}{2}
(1 - \bar{\gamma}_{5}),\qquad P^{+} = -
\frac{1}{2}\bar{\gamma}^{1}\bar{\gamma}^{0} = \frac{1}{2} (1 +
\bar{\gamma}_{5})\, ,
\end{equation}
which can be thought of as chirality projection operators and satisfy
\begin{equation}
P^{\pm}P^{\pm} = P^{\pm},\quad P^{\pm}P^{\mp} = 0,\quad P^{+}+P^{-} =
1\, .
\end{equation}

With the projection operators in (35), let us define the chiral
components
\begin{equation}
\psi_{\pm} = P^{\pm} \psi\, .
\end{equation}
The fermion part of the Lagrangian density for the Schwinger model, in
terms of these chiral components, takes the simple form
\begin{equation}
{\cal L} = -i \psi_{-}^{\dagger} \bar{\partial}_{1}\psi_{-} + i
\psi_{+}^{\dagger} (2\bar{\partial}_{0}+\bar{\partial}_{1})\psi_{+} -
e \psi_{-}^{\dagger}\psi_{-}\bar{A}_{1} + e \psi_{+}^{\dagger}\psi_{+}
(2\bar{A}_{0}+\bar{A}_{1})\, .
\end{equation}
It is clear from the form of the Lagrangian density in (38) that the
two fermion components are decoupled and  one of them, namely,
$\psi_{-}$  becomes
non-dynamical. As a result, it does not thermalize and the finite
temperature propagator for the fermions (in the real time formalism)
takes the form
\begin{equation}
iS_{-} (\bar{p}) = - \frac{i}{\bar{p}_{1}},\qquad iS_{+} (\bar{p}) = -
\frac{i}{2\bar{p}_{0}+\bar{p}_{1}} - 2\pi {\rm sgn} (\bar{p}_{1})
n_{\rm F} (|\bar{p}_{0}|) \delta (2\bar{p}_{0}+\bar{p}_{1})\, ,
\end{equation}
where ``sgn'' stands for the sign function and $n_{\rm F}$ represents the
fermion distribution function
\begin{equation}
n_{\rm F} (x) = \frac{1}{e^{\beta x} + 1}\, .
\end{equation}
The non-thermalization of the non-dynamical component is particularly
easy to see in the imaginary time formalism.

Given the fermion propagator in (39), we can now calculate the thermal
correction to the anomaly (coming from the fermion loop). There are
two contributions - one linear and the other quadratic in the
distribution function. The linera term in $n_{\rm F}$ has the form
(after factoring out the Lorentz structure)
\begin{equation}
4ie^{2}\pi \int \frac{\mathrm{d}^{2}\bar{k}}{(2\pi)^{2}}\, {\rm sgn}
(\bar{k}_{1}) n_{\rm F} (|\bar{k}_{0}|) \delta (2\bar{k}_{0} +
\bar{k}_{1}) = 0\, ,
\end{equation}
since the integrand is odd. The term quadratic in $n_{\rm F}$ vanishes
algebraically because of the identity
\begin{equation}
(2\bar{p}_{0}+\bar{p}_{1}) \delta (2\bar{k}_{0}+\bar{k}_{1}) \delta
  (2(\bar{k}_{0}+\bar{p}_{0}) + \bar{k}_{1}+\bar{p}_{1}) = 0\, ,
\end{equation}
so that there is no thermal correction to the anomaly as is the case
in conventionally quantized theories.

However, when one calculates the thermal corrections to the
self-energy (as well as higher point functions), they disagree with
the conventional calculations off-shell. Namely, on the mass-shell,
such thermal corections vanish thus coinciding with the conventional
results. However, off-shell, the thermal amplitudes in the light-front have
contribution only from one of the chiral components while in the
conventional calculation, both the chirality components thermalize and
contribute, leading to a different structure.

The Schwinger model is a simple theory where even bound state
questions can be studied [6]. This is best studied in the bosonized form
of the theory which corresponds to a free massive boson with the mass
related to the anomaly in the Schwinger model. Since the anomaly is
unaffected by temperature, it follows that the bound state equation as
well as the solution are independent of temperature. Finally, one can
also calculate the fermion condensate in the theory which is more
conveniently calculated using bosonization. Here, the temperature
dependent condensate takes the forms
\begin{eqnarray}
\langle \bar{\psi} \psi\rangle_{T} &\rightarrow & \langle
\bar{\psi}\psi\rangle_{T=0} \left(1 - \sqrt{\frac{2\pi T}{m_{\rm
      ph}}}\,e^{-\frac{m_{\rm ph}}{T}}\right),\qquad {\rm for}\; T {\rm
  small},\nonumber\\
\noalign{\vskip 4pt}%
 \bar{\psi}\psi\rangle_{T} & \approx &  - 2T\, e^{-\frac{\pi T}{m_{\rm
       ph}}},\qquad {\rm for}\; T {\rm
  large}\, .
\end{eqnarray}
This coincides with the results calculated with conventional
quantization [7].

\section{CONCLUSION}

In summary, we have shown in this talk how the naive generalization of
the density matrix to light-front field theories runs into serious
problems. We have identified the source of the difficulty to be the
absence of a rest frame for the heat bath on the light-front. We have
proposed an alternate generalization of the density matrix [2] which
is free from the divergence problems and has a natural meaning [3]
when analyzed systematically as a density matrix in a transformed
coordinate system. We have applied this formalism to the Schwinger
model and studied both perturbative as well as bound state
questions [4]. We have shown that the physical results coincide with the
ones calculated in conventional quantization. However, the structures
of the amplitudes off-shell do not quite coincide.

\vskip1cm
\noindent

\section*{ACKNOWLEDGEMENTS}
It is a pleasure to thank Simon Dalley and James Stirling for the
wonderful atmosphere at the LC03 conference. This work was supported
in part by USDOE Grant Number DE-FG 02-91ER40685.

\end{document}